\documentclass[]{aa}
\usepackage{graphicx}
\usepackage{natbib}
 
\bibpunct{(}{)}{;}{a}{}{,}     % A&A citation style
\newcommand{\micron}{$\mu$m}

\title{Spatial separation of small and large grains in the
transitional disk around the young star IRS~48\thanks{Based on
observations obtained at the European Southern Observatory, Paranal,
Chile, within the observing program 075.C-0211, and observations
obtained with the WHT operated on the island of La Palma by the Isaac
Newton Group.}}  \titlerunning{Spatial separation of grains in the disk
around IRS~48}

\author{
V.C. Geers \inst{1}
\and K.M. Pontoppidan\inst{2}
\and E.F. van Dishoeck\inst{1}
\and C.P. Dullemond\inst{3}
\and J.-C. Augereau\inst{4}
\and B. Mer\'{\i}n\inst{1,5}
\and I. Oliveira\inst{1}
\and J. W. Pel\inst{6}
}

\offprints{V.\,C. Geers,\\ \email{vcgeers@strw.leidenuniv.nl}}

\institute{Leiden Observatory, P.O. Box 9513, 2300 RA Leiden, The Netherlands
\and Hubble Fellow, Division of GPS, Mail Code 150-21, California Institute of Technology, Pasadena, CA 91125, USA
\and Max-Plank-Institut f\"{u}r Astronomie, Koenigstuhl 17, 69117 Heidelberg, Germany
\and Laboratoire d'Astrophysique de l'Observatoire de Grenoble, B.P. 53, 38041 Grenoble Cedex 9, France
\and Research and Scientific Support Department (ESTEC/ESA), Keplerlaan 1, 2200 AG Noordwijk, The Netherlands
\and Kapteyn Astronomical Institute, Landleven 12, 9747 AD Groningen, The Netherlands
}
\authorrunning{Geers et al.}

\date{Received 22 March 2007; Accepted 28 April 2007}
\abstract
{}
{We present spatially resolved mid-infrared images of the disk surrounding the young star \object{IRS~48} in the Ophiuchus cloud complex.
The disk exhibits a ring-like structure at 18.7\,$\mu$m, and is dominated by very strong emission from polycyclic aromatic hydrocarbons at shorter wavelengths. This allows a detailed study of the relative distributions of small and large dust grains.}
{Images of \object{IRS 48} in 5 mid-infrared bands from 8.6 to 18.7\,$\mu$m as well as a low resolution N-band spectrum are obtained with VLT-VISIR. Optical spectroscopy is used to determine the spectral type of the central star and to measure the
strength of the H$\alpha$ line.}
{The 18.7\,$\mu$m ring peaks at a diameter of 110\,AU, with a gap of $\sim 60$\,AU. The shape of the ring is consistent with an inclination of $i=48\degr\pm8\degr$. In contrast, the 7.5-13\,$\mu$m PAH emission bands are centered on the source and appear to fill the gap within the ring. The measured PAH line strengths are 10--100x stronger than those typically measured for young M0 stars and can only be explained with a high PAH abundance  and/or strong excess optical/UV emission.
The morphology of the images, combined with the absence of a silicate emission feature, imply that the inner disk has been cleared of micron-sized dust but with a significant population of PAHs remaining. We argue that the gap can be due to grain growth and settling or to clearing by an unseen planetary or low-mass companion. IRS 48 may represent a short-lived transitional phase from a classical to a weak-line T Tauri star.}
{}


\begin{document}
\maketitle

\section{Introduction}
The number of circumstellar dust disks detected around young stars has
increased dramatically over the last decades thanks to a variety of
ground- and space-based observations. Studies of the Spectral Energy
Distributions (SEDs) of young stellar objects at various ages indicate how
these disks evolve from optically thick, massive gas-rich disks to the
more optically thin, tenuous gas-poor disks by a combination of gas accretion, grain growth, planet formation and photo-evaporation of the gas. 

Only very few spatially resolved mid-infrared images have been presented.
Some tenuous disks around 5-15 Myr stars show evidence for gap
formation and spiral arm structures \citep[e.g.][]{jay98,aug99,liu04}, and
this has been interpreted as evidence for the presence of forming
planets clearing out a ring of dust and gas. For younger disks around
$\sim 1$\,Myr old stars, there is still very little direct evidence for
the formation of gaps \citep{fuj06}. 

An obvious step toward the formation of planetesimals in
 disks is the coagulation and growth of the sub-micron sized
 grains accreted from the proto-stellar envelope
 \citep{dom07}. Although the end results are plainly visible in our
 own planetary system, as well as in the emerging wealth of exo-solar
 planets, the details and mechanisms of dust evolution in
 proto-planetary disks are not well understood.  Very strong grain
 growth may lower the dust opacity enough to form an apparent gap in
 the disk at mid-infrared wavelengths, similar to those attributed to
 dust clearing by planets \citep{dal05,tan05}. A wide range of recent
 observational results indeed suggest that significant grain growth is
 a common occurence in disks \citep[e.g.][]{boe05,kes06,nat07}.

However, some observations complicate this picture. A subset of
protoplanetary disks shows strong emission features from extremely
small grains -- Polycyclic Aromatic Hydrocarbons (PAHs)
\citep{ack04,gee06}, even in disks with apparent gaps.
How do these disks fit into the general picture
of grain evolution? Are they evidence of grain size segregation,
leaving only the small grains in the upper layers of the disk?  Is gas
still present in these regions?

In this letter, we present one of the first spatially resolved mid-infrared
images and spectroscopy of a disk around a young (few Myr) star with 
a rising mid-IR SED, IRS~48 in Ophiuchus \citep{wil89} (catalogued as 
WLY 2-48, 16 27 37.19,  -24 30 35.0 J2000), which is no longer surrounded 
by an envelope. This source shows
exceptionally strong PAH emission at 3.3 and 7.7--12.3\,\micron\ in
the inner part of the disk {\it as well as an apparent inner gap seen
at 18.7\,$\mu$m}. We will 
discuss the origin of the 18.7\,$\mu$m gap in the context of the PAH
images which show that the gap is not {\it cleared} of material.

\section{Observations of \object{IRS~48} and data reduction}
\label{sec:obs}
\begin{figure}
  \centering
  \includegraphics[width=\columnwidth]{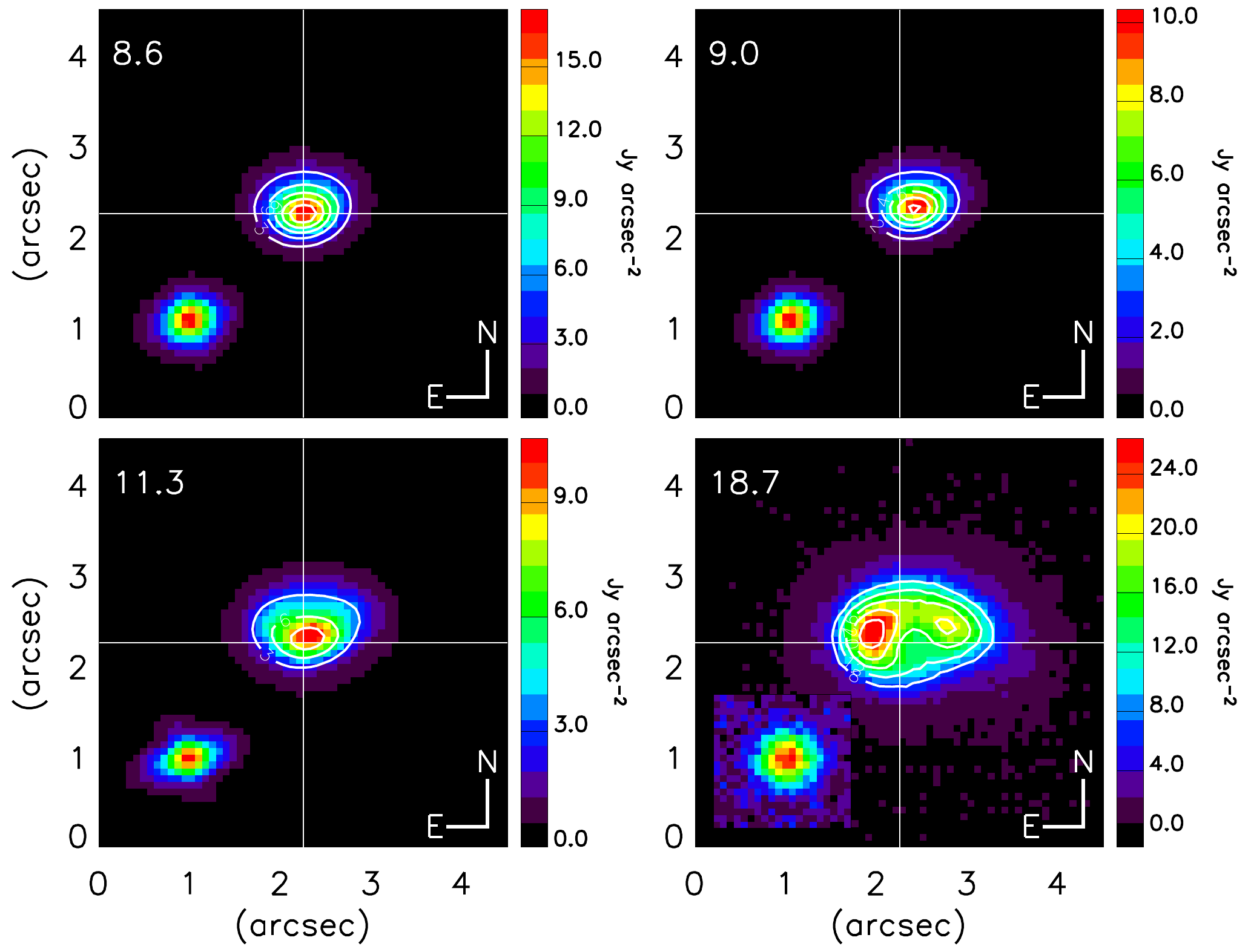}
  \caption{VISIR images of IRS~48, with PSF standard shown in inserts; the crosshair indicates the center of the emission at 8.6 $\mu$m. All images up to 12\,\micron\ are dominated by PAH emission. The 11.9\,\micron\ image (not shown) is similar to that at 11.3\,\micron. The contours are indicated for 3, 6, 9, 12, 15 (8.6\,\micron), 2, 4, 6, 8, 10 (9.0\,\micron), 3, 6, 9, 12 (11.3\,\micron) and 8, 12, 16, 20, 24 (18.7\,\micron) Jy arcsec$^{-2}$.
}
  \label{fig:visirimages}
\end{figure}
Images were obtained with VISIR on the Very Large Telescope in 5 mid-infrared bands at 8.6, 9.0, 11.3, 11.9 and 18.7\,\micron\ on June 9, 2005 (Fig.~\ref{fig:visirimages}).
\begin{figure}
  \centering
  \includegraphics[width=0.8\columnwidth]{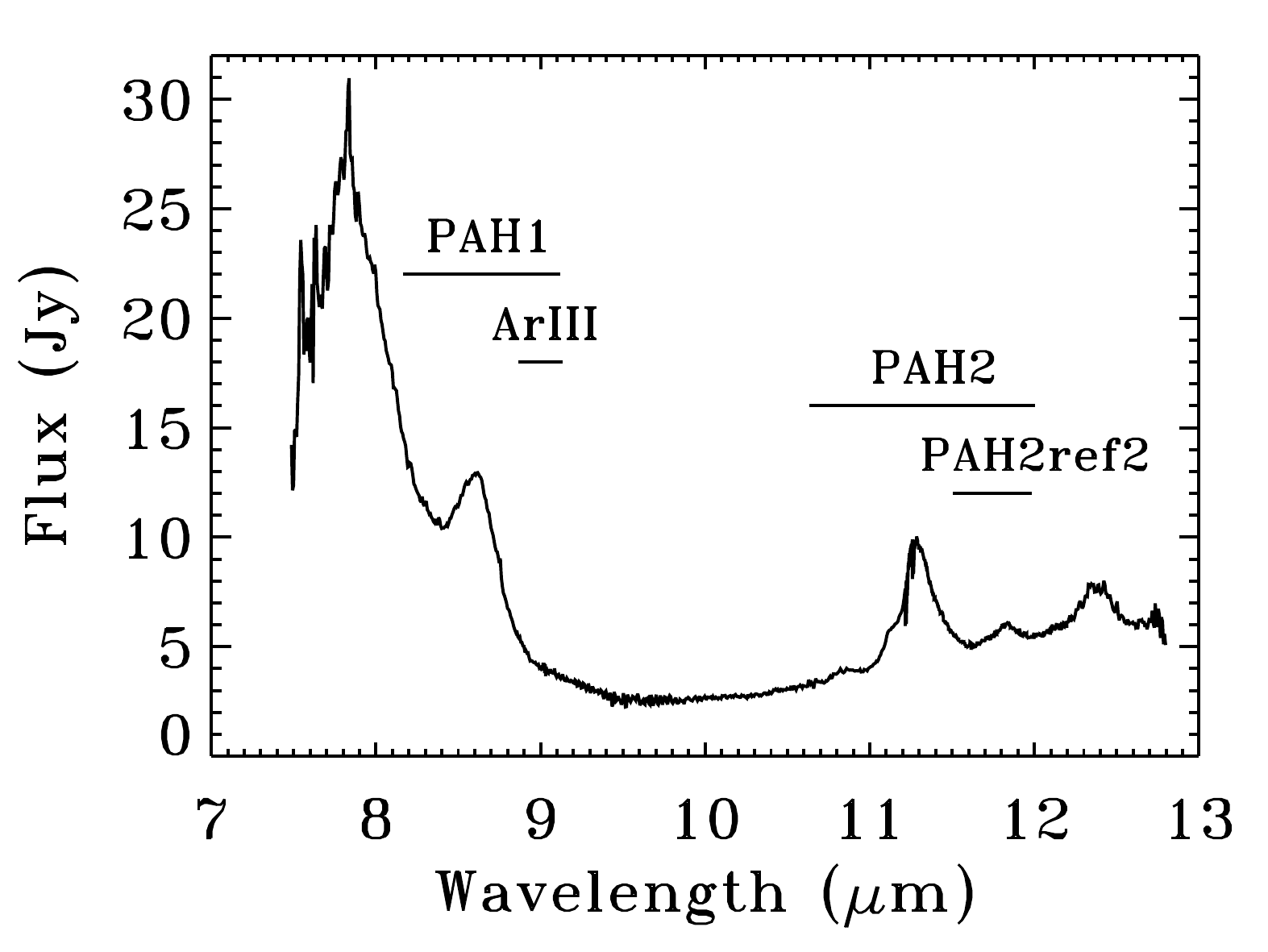}
  \caption{VLT-VISIR N-band spectrum of \object{IRS~48}. Spectral width of transmission curves
  of VISIR image filters are indicated.}
  \label{fig:visirspec}
\end{figure}
VISIR N-band spectroscopy was taken on June 12, 2005, in the low resolution settings at 8.8, 9.8, 11.4 and 12.2\,\micron, with a typical resolving power of $\lambda/\Delta\lambda \sim 300$ (Fig.~\ref{fig:visirspec}). The telescope was operated using a standard chop-nod scheme with chop-throws of 10$''$.  The data were reduced using a combination of the ESO pipeline (v.\,1.3.7) and our own IDL routines. 

\begin{figure}
  \centering
  \includegraphics[width=\columnwidth]{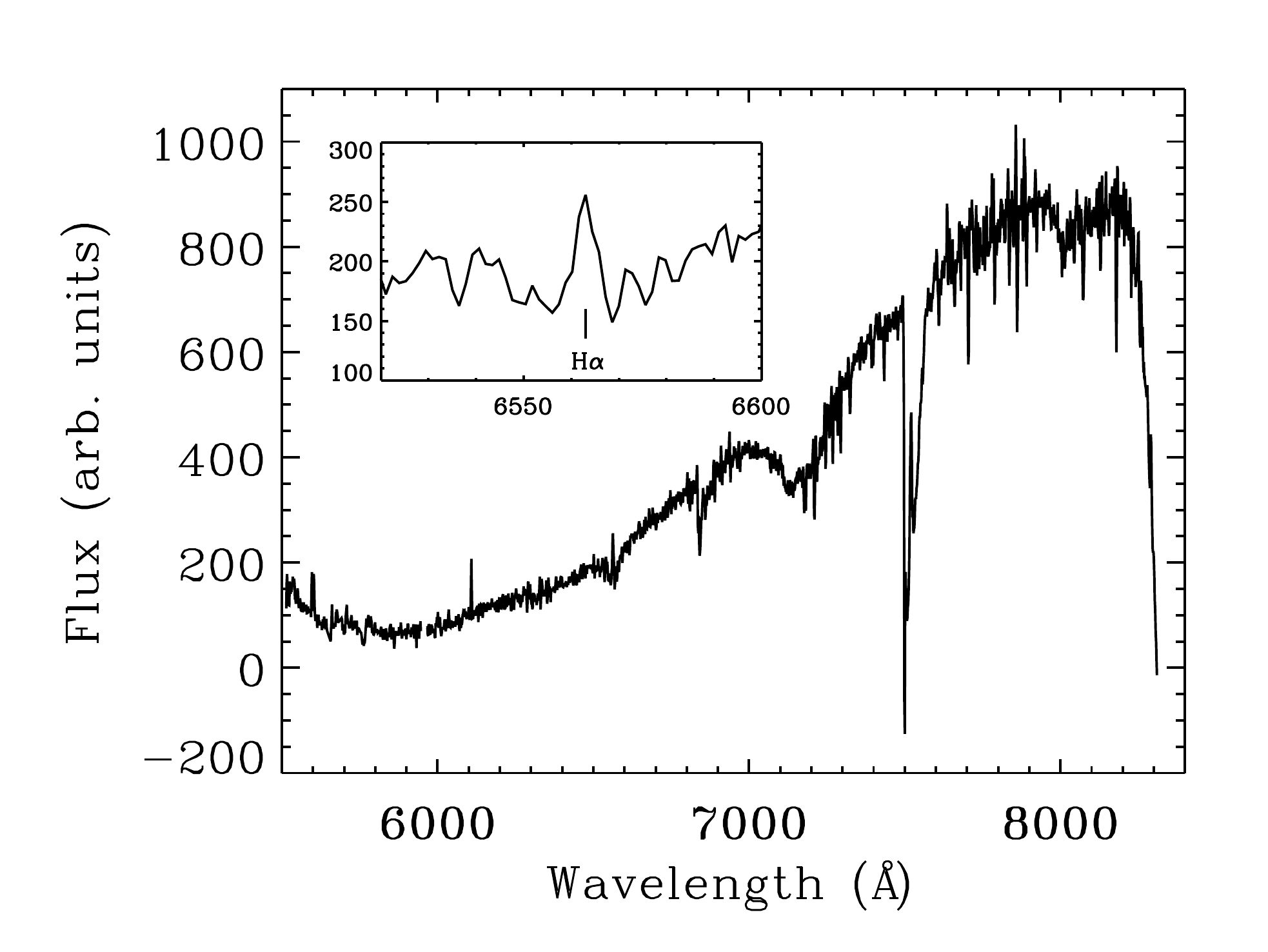}
  \caption{WHT-WYFFOS optical spectrum of IRS~48. The inset shows a blow-up of
  the H$\alpha$ line.}
  \label{fig:opt}
\end{figure}

A WYFFOS optical spectrum was obtained at the William Herschel
Telescope (WHT) on La Palma on May 4, 2006, with a resolving power of
$R \sim 1600$ (Fig.~\ref{fig:opt}).
It indicates a M0 spectral type ($T_{\rm eff}$ = 3800 K) with an error of less than 2 spectral
subtypes. $A_v$ = 7 $\pm$ 1 mag is derived from the optical spectrum as in Oliveira et al. (in prep.).
H$\alpha$ is detected with an
Equivalent Width (EW) of 5.9 \AA, which classifies \object{IRS~48} as a
weak-line T Tauri star. Our derived spectral type is much later than
that of \citet{luh99}, who found it to be earlier than F3 based on a
K-band spectrum (taken sometime between July 1994 and June 1996). 
This discrepancy remains to be understood and will
be discussed in \S \ref{ssec:fuori}. 

Photometry used to construct the SED (Fig.~\ref{fig:sed}) includes NOMAD \citep{zac04}, USNO-B \citep{mon03}, 2MASS \citep{skr06}, ISOCAM \citep{bon01}, Spitzer Space Telescope IRAC and MIPS (3.6, 4.5, 5.8, 8.0, 24, 70\,\micron) from the c2d legacy program database \citep{eva03}
and IRAM 1.3 mm \citep{and94}.

\section{Results}
The circumstellar disk is spatially resolved in all 5 VISIR images,
shown in Fig.~\ref{fig:visirimages}.  An extent of 1.2--1.9$''$ is derived along the semi-major axis, corresponding to 150--230 AU diameter; the actual extent increases with wavelength. \citet{duc05} measured a proper motion consistent with the surrounding Oph sources placing \object{IRS~48}
firmly at a distance of 125 pc \citep{geu89}. All 5 bands are resolved with respect to the standard star PSF along the east-west direction, showing elongated surface brightness profiles consistent with a circumstellar disk. We derive an inclination of $i=48\degr\pm8$\degr\ from the semi-major and minor axes of ellipsoidal contours fitted to the 18.7\,\micron\ image, and a position angle of 98 $\pm$ 3\degr\ East of North. 

The most striking result is that the 18.7\,\micron\ Q-band image shows an asymmetry in the surface
brightness and a gap in the center. We interpret the apparent shape as being due to an inclined ring-shaped disk. No point source is seen
within this gap. Its diameter as measured from the inner edges of the ring along the semi-major axis equals 0.5$''$ or $\sim 60$ AU, corresponding to a gap with a radius of $\sim 30$ AU. In contrast, the PAH emission at 8.6 and 11.3\,\micron\ is centrally peaked with resolved wings beyond a point-source, and it appears to originate from the apparent gap in the Q2 image. 
The center of the images shifts by $\sim 0.2''$, likely due to pointing error, as indicated by similar shifts in the standard star positions.
The PAH off-band filters at 9.0 and 11.9\,\micron\ both resemble the 8.6 and 11.3\,\micron\ images respectively. Given the strength of the PAHs (see Fig.~\ref{fig:visirspec}), it is assumed that both off-band filters also probe PAH emission or very small grains.

The VISIR N-band spectrum (Fig.~\ref{fig:visirspec}) shows clear PAH
features at 8.6, 10.8, 11.3, 11.9 and 12.5 $\mu$m. 
Our 11.3 $\mu$m line
flux is $2.5\times 10^{-14}$ W m$^{-2}$. The PAH features around this
M0 star are as strong as the strongest features observed around Herbig
Ae/Be stars (Acke et al. 2004), and make it the latest type young star
with detected PAHs. No silicate emission feature at 9.7\,\micron\ is
detected.  A VLT-ISAAC L-band (2.8--4.2\,\micron) spectrum, obtained as part of the ice survey by  \citet{dis03}, taken on 2002, May 5 (sample and reduction described in \citealt{pon03}), shows the presence of a strong 3.3\,\micron\ PAH
feature and the 4.05\,\micron\ Br$\alpha$ line (included in Fig.~\ref{fig:sed}).

\section{Discussion}
\subsection{Gap in the disk}
\begin{figure}
  \centering
  \includegraphics[width=0.8\columnwidth]{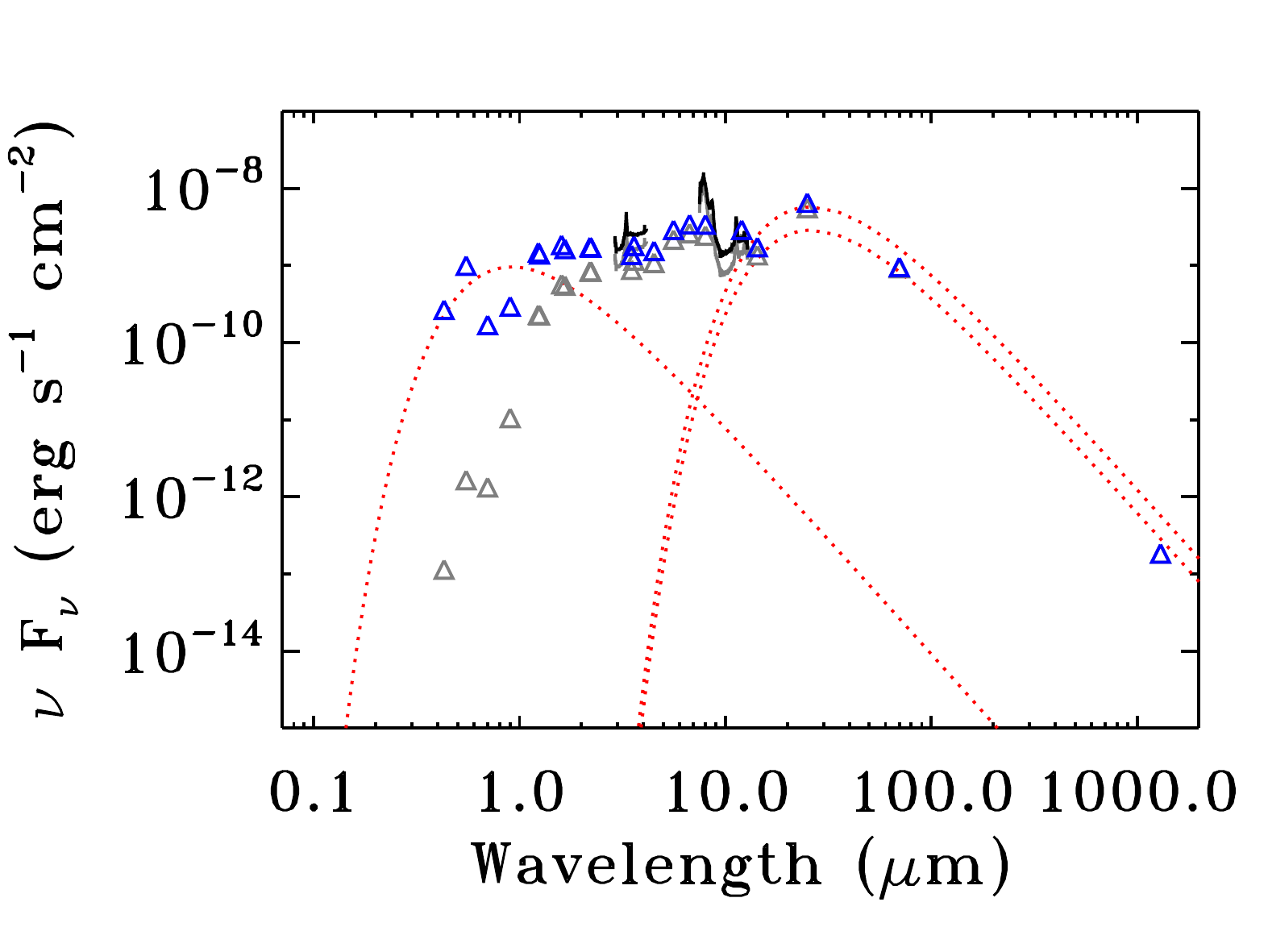}
  \caption{The SED of \object{IRS~48} based on the
  photometry listed in \S \ref{sec:obs}. Grey triangles are literature photometry, blue triangles are dereddened, with $A_{\mathrm{v}} = 7$; black solid lines are ISAAC L-band and VISIR N-band spectra; orange dotted lines show 3 blackbodies, at $T_{\mathrm{eff}}$ = 4000 K (0.63 L$_{\odot}$) and 140 K, the latter scaled to 1.9 and 3.8 L$_{\odot}$. }
  \label{fig:sed}
\end{figure}
Gaps in disks around T Tauri stars have been inferred for a few
sources (\citealt{cal05,aur06}; Brown et al.\ submitted) but entirely on the
basis of the SED, not spatially resolved images such as presented
here. The 18.7\,\micron\ image suggests a gap with a radius of 30 AU in
the dust disk. However, the PAH-band images show that this gap cannot be entirely devoid of small particles. The near-infrared excess at 1--3\,\micron\ also suggests that some hot dust
still exists in a small ring of material or a puffed-up inner rim in the inner few AU, i.e.,
the disk shows a gap and not a hole. 
Interestingly, the SED of \object{IRS~48} (Fig.~\ref{fig:sed}) does not reveal the presence of a gap, possibly due to masking of the dip at $\sim 5$--15\,\micron\ by the strong PAH features and other types of very small grains (VSGs) present in the gap. Quantum-heated PAHs and very small grains reach a much higher average temperature than thermal grains, radiating strongly in distinct PAH features and continuum at 5--15\,\micron, while being weaker at 20\,\micron.

Gaps in disks have been interpreted in the context of photo-evaporation 
of gas and small dust grains by the central star. This mechanism is excluded 
here since it should create a similar gap in the PAH filter images. 
Another way to make an apparent gap is to lower the
dust opacity by grain growth, effectively removing the silicate dust
particles responsible for the continuum emission between a few and
$\sim 15$\,\micron. Theoretical models show that grain growth occurs on
short time scales, with the shortest growth times in the inner parts
of the disk \citep{wei97,dul05}. Furthermore, larger grains settle to
the mid-plane faster than smaller grains, creating a strong size
segregation in the vertical direction, which would explain the strong
feature/continuum ratio of the PAH features (Dullemond et al.\ subm.). 
The lack of a significant silicate feature at 9.7\,\micron\ supports 
the idea that in the inner disk most of the dust is in larger micron-sized grains. 
However, this scenario has difficulties explaining the presence of a sharply defined ring structure.

A third way of gap formation is by the clearing out
of gas and dust by a planet forming inside the disk \citep[e.g.][]{kla01} or a low-mass companion. 
Planets are expected to clear out larger ($>$100\,\micron) grains much 
more rapidly than gas and small particles by tidal interaction and can cause 
sharp edges in images \citep{paa04,qui04}. Particularly, \citet{ric06} predict that larger dust grains are filtered out at the outer edge, presumably leading to a build-up of large dust grains and a clearly defined ring of material at the outer edge, while smaller grains (e.g.\ PAHs) continue to accrete inwards along with the gas, which would explain the continued presence of PAH emission in the gap. Interferometric observations at millimeter wavelengths to constrain the distribution of even larger (mm-sized) particles can further test these scenarios.

\subsection{Source of central luminosity}
\label{ssec:fuori}
The SED of \object{IRS~48} is peculiar for several reasons. It shows a very strong bump at 25\,\micron\ which appears to be consistent with a single temperature blackbody of $\sim 140$ K. This
emission bump includes 18\,\micron\ and should be associated with the ``ring'' 
seen in the Q2-band image.  The
apparent luminosity from the 25\,\micron\ bump is $\sim 1.9$--3.8
L$_{\odot}$.
If this ring is assumed to be the sole source of the bump and if we 
assume the ring to be an optically thick annulus with a 
covering fraction of at most 0.2 with respect to the star, it follows that the central source
should have a luminosity of at least 10--20 L$_{\odot}$, which is more
than 20 times stronger than expected for a 1 Myr M0 star.

External heating by nearby bright sources can be excluded from our own VISIR, Spitzer IRAC 
and ESO archive NACO images, as well as recent multiplicity surveys \citep{hai04}. The possibility of a close ($<$ 175 AU) binary with a higher mass early type star can be excluded for lack of spectral features associated with early type stars in our optical spectrum.  

The most likely scenario is the presence of excess UV emission, which is absorbed by a local foreground layer of material, such as a strongly inclined disk with an inner rim that is puffed up due to temporary accretion events, or absorption by the flared outer disk. A fully edge-on disk is considered unlikely; it could produce the observed SED, but it would be inconsistent with the Q2-band image. 

One example of excess luminosity is through accretion. This would require an accretion luminosity much larger than the star itself.
No X-ray emission has been detected toward
this object by Chandra nor XMM-Newton (Grosso, priv.\ comm.)  and both
H$\alpha$ and Pf$\beta$ are relatively weak, which argues against
strong accretion.

An alternative explanation might be that this disk has recently
undergone a FU Orionis outburst. During a major FU Ori-type accretion
event, the temperature of the disk increases over a span of 1--10 years
from a few hundred to a few thousand K, and will dominate the spectrum
even for optical wavelengths \citep{har01}. The apparent
spectral type can change by several types, from late K-M to F-G
type. This may explain the discrepancy between the $<$F3 spectral type
determination observed between 1994 and 1996 and the M0
type determination in 2006. Heating of the disk would help the PAHs in two ways: any PAHs
trapped in ices would be evaporated boosting the PAH abundance and the
PAHs could be thermally excited.

\subsection{PAH feature strength}
\object{IRS~48} is exceptional in its very prominent PAH features, both in
strength and feature-over-continuum ratios, 100--1000 times stronger than recent model predictions for M0 stars \citep{gee06}.  
Comparing our measured 11.3\,\micron\ line strength with their Fig.~9, we find that it would be consistent with the radiation field of a $\sim 6000$ K central star, in a disk model with a PAH abundance of 5$\times$10$^{-5}$ with respect to hydrogen, typical of the general interstellar medium but higher than inferred for disks. Assuming this abundance, 
the strong PAH features would be consistent with excess optical/UV 
luminosity at a level corresponding to several hundred times the 
average interstellar radiation field at a radius of 100 AU, which is 
consistent with recent optical/UV strengths of T Tauri stars inferred by \citet{ber03}.

The case of \object{IRS~48} suggests that the combination of high feature-over-continuum 
PAH bands and the absence of silicate features together with a rising 
SED at $\lambda > 12$\,\micron\ is also an indicator for the presence of 
gap formation through grain growth, which is not
revealed in broadband SEDs. Mid-infrared spectroscopy and high-spatial
resolution narrow-band imaging with 8-m class telescopes, combined with millimeter interferometric imaging, of a much
larger sample of objects is crucial to determine whether \object{IRS~48} is
just a peculiar object or whether it forms part of a new class of
transitional disks.

\begin{acknowledgements}
We thank A.\ Smette for crucial help in obtaining the VISIR data and N.\ Grosso for providing X-ray information. KMP is supported by NASA through Hubble Fellowship grant 01201.01 awarded by the STScI, which is operated by the AURA, for NASA, under contract NAS 5-26555. Astrochemistry in Leiden is supported by a Spinoza grant from the Netherlands Organization for Scientific Research (NWO).
\end{acknowledgements}

%###############
%\bibliographystyle{aa}
%\bibliography{7524}

%###############

 \end{document}